\documentclass[12pt,preprint]{aastex}

\slugcomment{Submitted to {\it The Astrophysical Journal (Letters)}}

\shorttitle{Serendipitous X-ray Detection of NGC~5315}
\shortauthors{Kastner et al.}

\begin{document}


\title{Serendipitous {\it Chandra} X-ray Detection of a Hot Bubble
  within the Planetary Nebula NGC~5315}


\author{Joel H. Kastner\footnote{ Visiting Astronomer, Laboratoire
  d'Astrophysique de Grenoble, Universit\'e Joseph Fourier --- CNRS,
  BP 53, 38041 Grenoble Cedex, France } \and Rodolfo Montez, Jr.} 
\affil{Chester F. Carlson Center for Imaging
  Science, Rochester Institute of Technology, 54 Lomb Memorial Dr.,
  Rochester, NY 14623 } \email{jhk@cis.rit.edu, rxm9447@cis.rit.edu} 


\author{Bruce Balick}
\affil{Department of Astronomy, University of Washington, Seattle, WA 98195--1580}
\email{balick@astro.washington.edu}

\and

\author{Orsola De Marco}
\affil{Department of Astrophysics, American Museum of Natural History, Central Park West at 79th Street, New York, NY 10024}
\email{orsola@amnh.org}





\begin{abstract}
  We report the serendipitous detection of the planetary nebula
  NGC~5315 by the {\it Chandra} X-ray Observatory. The {\it Chandra}
  imaging spectroscopy results indicate that the X-rays from this PN,
  which harbors a Wolf-Rayet (WR) central star, emanate from a $T_X
  \sim 2.5\times10^6$ K plasma generated via the same wind-wind
  collisions that have cleared a compact ($\sim8000$ AU radius)
  central cavity within the nebula. The inferred X-ray luminosity of
  NGC~5315 is $\sim2.5\times10^{32}$ erg s$^{-1}$ (0.3-2.0 keV),
  placing this object among the most luminous such ``hot bubble''
  X-ray sources yet detected within PNe. With the X-ray detection of
  NGC~5315, objects with WR-type central stars now constitute a clear
  majority of known examples of diffuse X-ray sources among PNe; all
  such ``hot bubble'' PN X-ray sources display well-defined,
  quasi-continuous optical rims. We therefore assert 
  that X-ray-luminous hot bubbles are characteristic of
  young PNe with large central star wind kinetic energies and closed
  bubble morphologies. However, the evidence at hand also suggests
  that processes such as wind and bubble temporal evolution, as well
  as heat conduction and/or mixing of hot bubble and nebular gas,
  ultimately govern the luminosity and temperature of superheated
  plasma within PNe.
\end{abstract}


\section{Introduction}

Imaging spectroscopy of planetary nebulae (PNe) by the {\it
Chandra} and {\it XMM-Newton} X-ray observatories has
yielded a steadily accumulating body of compelling
observational evidence for ``hot bubbles'' within certain
PNe. Such a hot bubble may be produced as the central star
makes the transition from post-asymptotic giant branch
(post-AGB) star to white dwarf, following an evolutionary
track of increasing $T_\star$ at constant $L_\star$,
followed by decreasing $L_\star$ and $T_\star$. In this
phase the star produces very fast and energetic winds (with
speeds $\sim1000$ km s$^{-1}$ and mass loss rates
$\stackrel{>}{\sim}10^{-7}$ $M_\odot$ yr$^{-1}$).  When such
a fast wind collides with ambient (previously ejected AGB)
gas, it is shocked and superheated (e.g., Zhekov \&
Perinotto 1996). The shocked fast wind forms an
overpressured bubble that accelerates outwards and displaces
the ambient (visible, nebular) gas as it grows (Kwok,
Purton, \& Fitzgerald 1978). The supersonic growth of the
bubble plows the displaced older material into a rim of
dense gas which, when projected on the sky, is seen as
a thin molecular, dust, and/or ionized structure that traces the
bubble's perimeter.

So far, nine of $\sim25$ PNe targeted by {\it Chandra} or
{\it XMM} have been detected as diffuse X-ray sources
(Kastner et al.\ 2000, 2001, 2003; Chu et al.\ 2001;
Guerrero et al.\ 2002, 2005; Sahai et al.\ 2003; Montez et
al.\ 2005; Gruendl et al.\ 2006). In most of these nebulae,
the X-ray-emitting region is fully contained within bright
optical rims or bubbles, as predicted by the preceding ``hot
bubble'' scenario. Furthermore, many of the detected objects
harbor central stars that are of Wolf-Rayet (WR) type (i.e.,
``[WC]'' or ``[WO]'' stars) and/or display optical
spectroscopic evidence for large mass-loss rates and wind
velocities. In this {\it Letter}, we report the
serendipitous detection by {\it Chandra} of X-ray emission
associated with an additional [WC] PN, NGC~5315, and we
present evidence that this emission also arises within a hot
bubble. The accidental detection of NGC~5315 underscores
certain trends that are emerging from X-ray detections of PN
hot bubbles over the past decade.

\section{X-ray Detection of NGC~5315}

\begin{figure}[htb]
\begin{center}
\includegraphics[scale=1.0,angle=0]{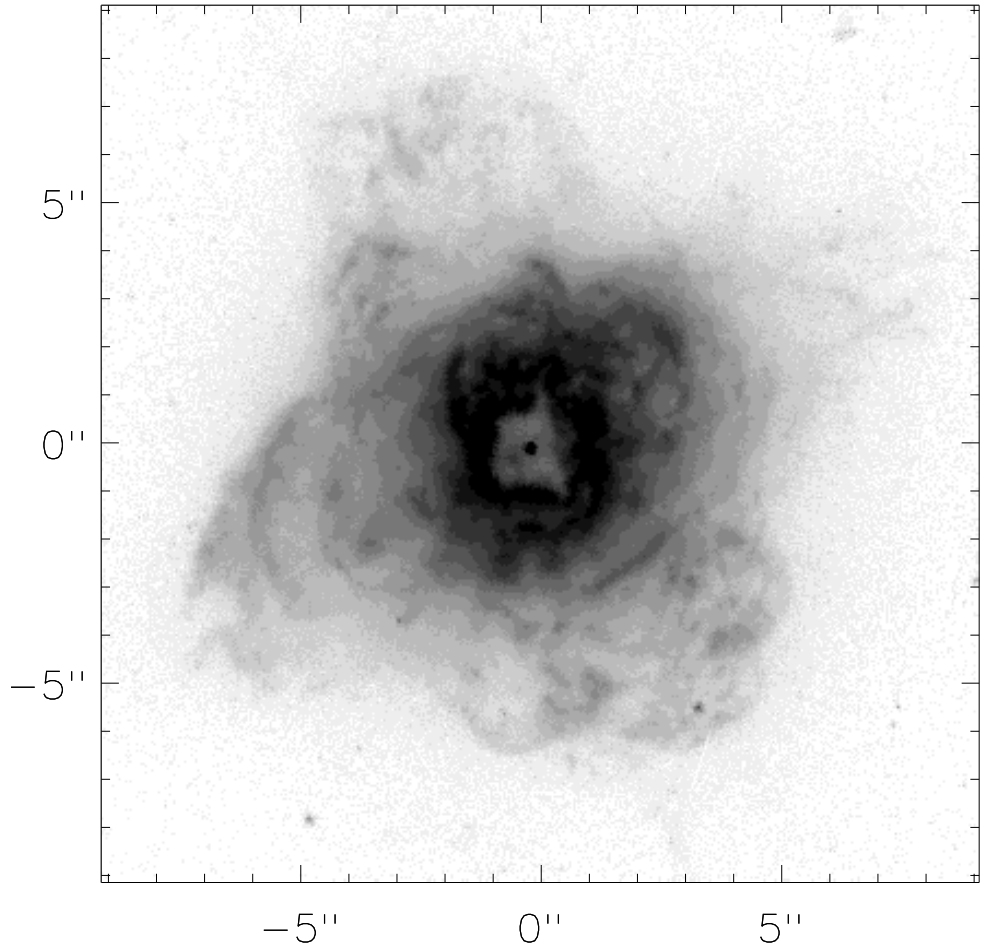} 
\end{center}
\caption{{\it Hubble Space Telescope} WFPC2 H$\alpha$ (F656N) image
  of NGC~5315 obtained 2007 February. North is up and east is to the left.}
\label{fig:HST}
\end{figure}

\begin{figure}[htb]
\begin{center}
\includegraphics[scale=0.7,angle=0]{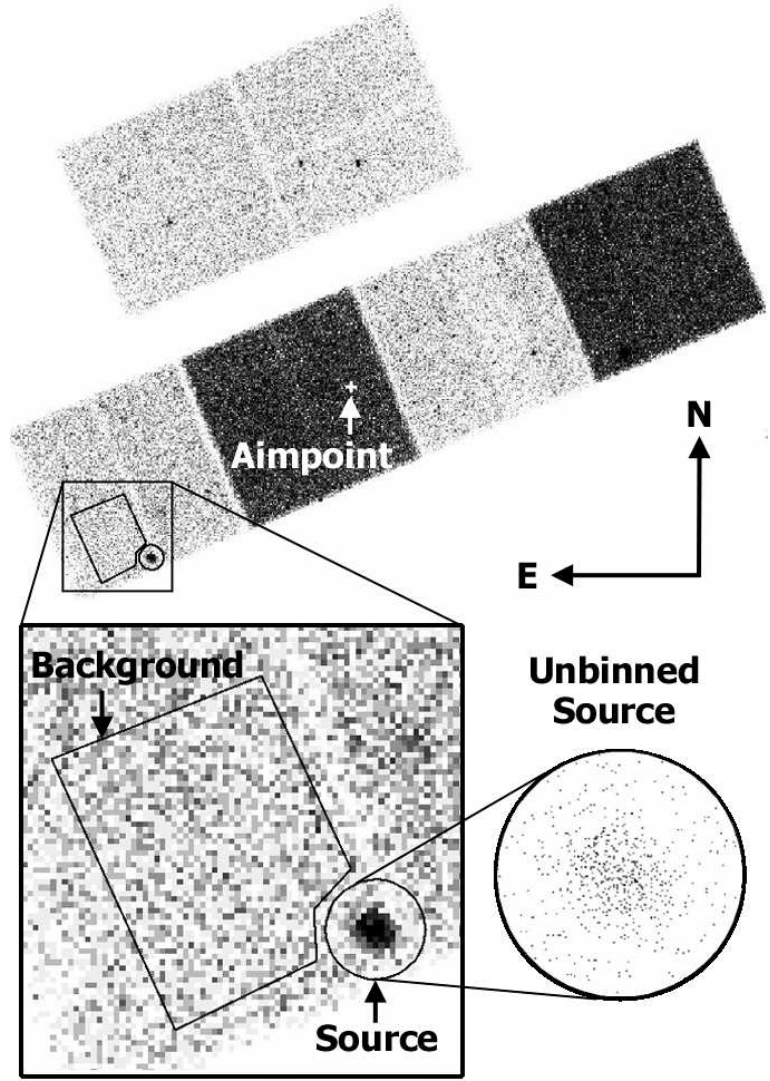}
\end{center}
\caption{Serendipitous Chandra/ACIS observation of NGC~5315. Displayed
  are rebinned ($3.9''\times3.9''$ pixel) images of the entire 6-CCD
  field (upper half of figure) and a smaller ($5'\times5'$) region
  that includes the NGC 5315 X-ray source (lower left), as well as an
  unbinned ($0.49''\times0.49''$ pixel) image of the ($34''$ radius)
  source spectral extraction region (lower right). The background
  spectral extraction region is also indicated.}
\label{fig:Chandra}
\end{figure}

In a Cycle 5 Chandra study, we observed two [WC] PNs, NGC 40
and Hen 2-99; we detected the former, but failed to detect
the latter (Montez et al.\ 2005).  Recently, while searching
the Chandra archives for targeted observations of PNs, one
of us (Montez) established that a {\it second} [WC] PN,
NGC~5315 (Fig.~\ref{fig:HST}), was present in the 29 ks
Chandra Advanced CCD Imaging Spectrometer (ACIS) observation
that targeted Hen 2-99 (OBSID 4480). The position of NGC~5315 lies just
within the field of view of the ACIS-S detector array, on a
CCD (the front-illuminated S4) that lies adjacent to the
prime imaging CCD (the back-illuminated S3). Examination of
the Chandra/ACIS image revealed an X-ray source at the
position of NGC~5315, $\sim12.5'$ off-axis
(Fig.~\ref{fig:Chandra}).

Due to the large off-axis angle and the fact that --- when
not used in conjunction with the high-energy transmission
gratings --- ACIS-S4 deviates significantly from the focal
surface of the Chandra mirror assembly, the NGC~5315 X-ray
source is subject to severe image aberrations. The Chandra
PSF semimajor axis is $\sim10-20''$ at such off-axis
angles\footnote{See e.g.
http://cxc.harvard.edu/ccw/proceedings/02\_proc/presentations/c\_allen/offaxis\_psf.pdf.gz},
without accounting for the displacement of the S4 detector
from the focal surface. Hence, apart from establishing that
an X-ray source is coincident with the position of NGC~5315,
no spatial information can be readily extracted from the
ACIS-S4 image.

\begin{figure}[htb]
\begin{center}
\includegraphics[scale=0.7,angle=-90]{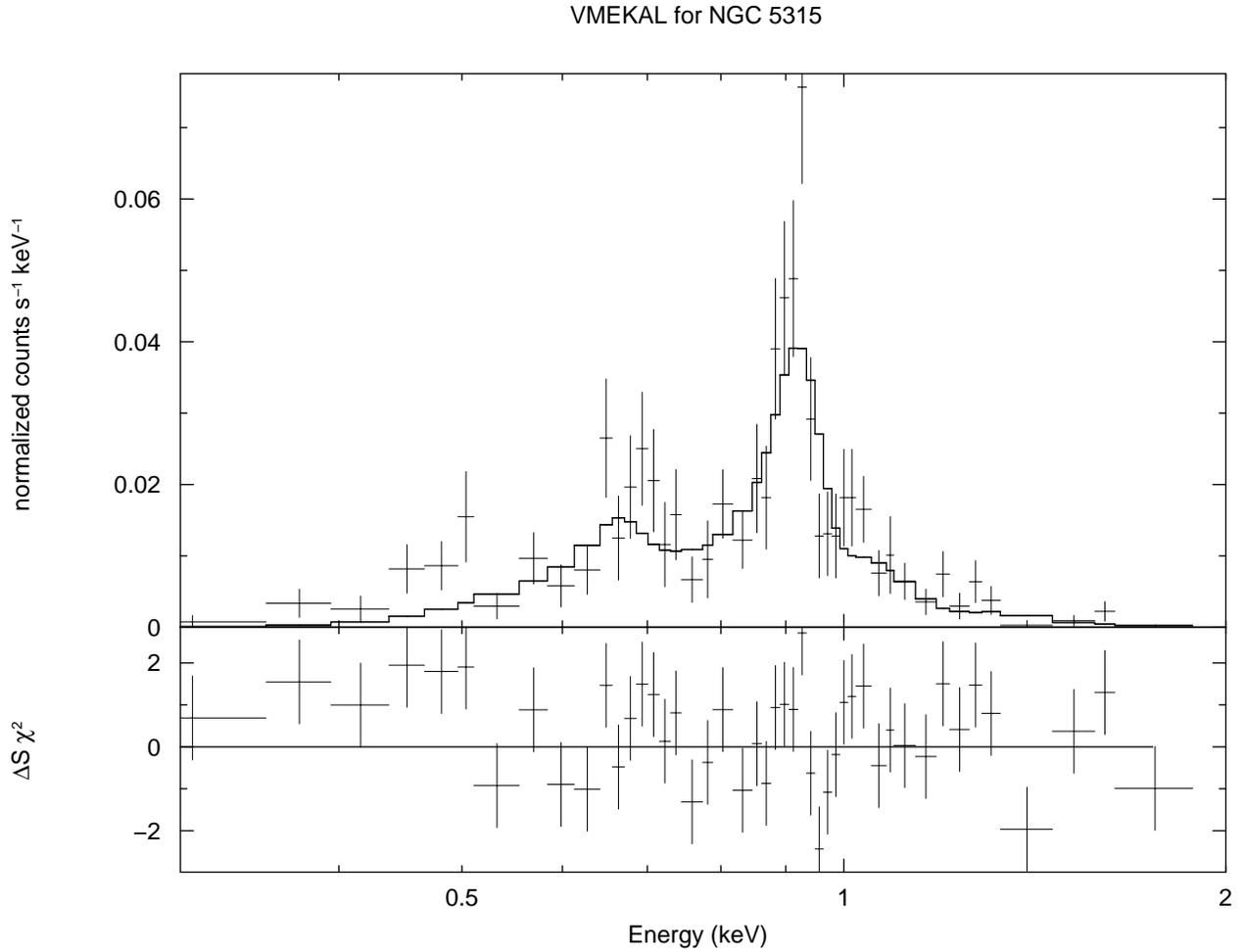}
\end{center}
\caption{ Chandra/ACIS spectrum of the X-ray
  source associated with NGC~5315 (crosses), with best-fit
  absorbed thermal plasma (VMEKAL) model spectrum
  overlaid. The fit residuals are indicated in the lower panel.}
\label{fig:NGC5315spec}
\end{figure}

We used version 3.4 of
CIAO\footnote{http://cxc.harvard.edu/ciao/} to extract
source and background spectra, spectral responses, and light
curves. The circular source extraction region radius was
34$''$ (Fig.~\ref{fig:Chandra}), which should encompass the
95\% encircled energy of a point source at the off-axis
angle corresponding to the position of NGC~5315.  Background
was extracted from a $\sim2.5'\times3.5'$ rectangular region
adjacent to the NGC~5315 source
(Fig.~\ref{fig:Chandra}).  The resulting,
background-subtracted spectrum displays prominent Ne {\sc
ix} emission at $\sim0.9$ keV and a blend of O {\sc vii} and
O {\sc viii} emission lines at $\sim0.65$ keV
(Fig.~\ref{fig:NGC5315spec}).The background-subtracted
count rate of the source is $12.4 \pm 0.7$ ks$^{-1}$. The
light curve (not shown) reveals no significant variation in
this count rate during the 28.7 ks exposure.

\section{Spectral Modeling}

We used version 12.3.0 of
XSPEC\footnote{http://heasarc.nasa.gov/docs/xanadu/xspec/}
(Arnaud 1996) to fit absorbed thermal plasma emission models
to the NGC~5315 X-ray source
(Fig.~\ref{fig:NGC5315spec}). The model absorbing column was
fixed at $\log{N_H} ({\rm cm}^{-2}) = 21.36$, corresponding
to the measured extinction toward NGC~5315 ($A_V \approx
1.3$; Peimbert et al.\ 2004; Pottasch et al.\ 2002; Cahn et
al.\ 1992). Applying the VMEKAL variable-abundance plasma
emission model (Liedahl et al.\ 1995 and references
therein), we find the best-fit plasma temperature is $T_x =
2.6\times10^6$ K ($\pm10$\%) and the inferred intrinsic
X-ray luminosity is $L_x \sim 2.6\times10^{32}$ erg s$^{-1}$
(0.3--2.0 keV) assuming a distance\footnote{Estimates of the
distance to NGC~5315 range from $\sim$1.3 kpc (Cahn et al.\
1992; Pottasch 1982) to 2.6 kpc (Gathier et al.\ 1986).} of
2.5 kpc (Marcolino et al.\ 2007). The plasma modeling
suggests Ne is enhanced and Fe is depleted; the best-fit
VMEKAL model abundances are $4.3 \pm 1.3$ and $0.6 \pm 0.3$,
respectively, relative to solar (Anders \& Grevesse 1989).

\section{Discussion}

Due to the very poor off-axis {\it Chandra}/ACIS-S image
quality at the position of NGC~5315, it is not possible to
ascertain from the ACIS image alone whether the X-rays trace
a hot bubble within this PN, emanate from the PN nucleus, or
are emitted by both the nebula and its central
star(s). However, the background-subtracted spectrum
(Fig.~\ref{fig:NGC5315spec}), the luminosity, and the
temporal behavior of the X-ray source associated with
NGC~5315 appear quite definitive regarding the origin of the
X-rays. The spectrum shows strong Ne~{\sc ix} line emission
as well as a blend of O~{\sc vii} and O~{\sc viii} lines,
with no evidence for Fe L-shell lines. Spectral modeling
indicates that the emission arises in a $\sim2.5$ MK thermal
plasma with enhanced Ne and depleted Fe.  These results and
the inferred source $L_X$ are very similar to those obtained
for the best-characterized diffuse X-ray PN, BD
+30$^\circ$3639 (Kastner et al.\ 2000, 2006; Maness et al.\
2003). Meanwhile, the Chandra/ACIS light curve shows no
evidence for variability, and the absorption-corrected X-ray
luminosity of NGC~5315 is at least an order of magnitude
larger than that of any unresolved PN core region detected
thus far (Guerrero et al.\ 2001; Kastner et al.\ 2003),
further supporting the interpretation that the X-rays arise
from an extended region within NGC~5315.

Assuming the X-rays from NGC~5315 indeed arise from its
compact ($\sim1''$ radius), sharply delineated central
cavity (Fig.~\ref{fig:HST}), the results reported here
establish NGC~5315 as one of the most luminous ``hot bubble'' X-ray
sources yet detected (Table 1). The detection of NGC~5315 by 
{\it Chandra} therefore underscores three significant trends that
have emerged from the X-ray observations of PNe obtained
thus far by {\it Chandra} and {\it XMM}:
\begin{enumerate}
\item Objects with WR-type (i.e., [WC], [WO], or WR(H)) central stars
  --- which display characteristically large wind velocities ($v_w$), large
  mass-loss rates ($\dot{M}$), and relatively low effective
  temperatures ($T_{\rm eff} \sim 30-60$ kK; e.g., Crowther et al.\
  1998) --- account for five of the seven detections of ``hot bubble''
  (as opposed to jet-excited) PN X-ray sources (Table 1). Given that
  $\stackrel{<}{\sim}10$\% of galactic PNe are known to harbor WR-type
  central stars (Gorny \& Stasinska 1995; Tylenda 1996), such PNe (and
  [WC] PNe in particular) clearly constitute a disportionately large
  fraction of objects that are established sources of luminous,
  diffuse X-ray emission. One of the two PNe in Table 1 that does not
  have a WR-type central star, NGC 7009, displays a very large central
  star wind velocity. The only X-ray nondetection among the [WC] PNe, the cool
  ([WC 9]) and very young Hen 2-99, appears to contain a bright,
  compact core, and hence may not yet have formed a central,
  wind-blown bubble (Montez et al.\ 2005).
\item All of the Table 1 objects --- not just the WR-type PNe ---
  feature central stars that are relatively cool and luminous ($T_{\rm
    eff} \stackrel{<}{\sim}80$ kK and $L \sim 3000-10000$ $L_\odot$;
  e.g., Mal'Kov 1997) confirming that PNe with X-ray-luminous bubbles
  are quite young (Soker \& Kastner 2003) and suggesting such PNe are
  generally the descendants of relatively massive
  progenitors. Specifically, comparison with theoretical post-AGB
  evolutionary tracks (e.g., Fig.\ 1 of Perrinoto et al.\ 2004)
  suggests post-AGB ages $\stackrel{<}{\sim} 3000$ yr (in general
  agreement with the dynamical ages of these PNe; see, e.g., Fig.\ 7
  of Akashi et al.\ 2006) and progenitor masses in the range $1-5$
  $M_\odot$.
\item In all cases in which diffuse X-ray emission is detected, the
  optical/IR structures that enclose the regions of diffuse X-rays are
  clearly defined, and these structures generally display thin,
  bright, uninterrupted edges (or ``rims'') surrounding a cavity of
  lower surface brightness that is coincident with the extended X-ray
  emission (see also Gruendl et al.\ 2006).
\end{enumerate}

\begin{figure}[htb]
\begin{center}
\includegraphics[scale=0.7,angle=90]{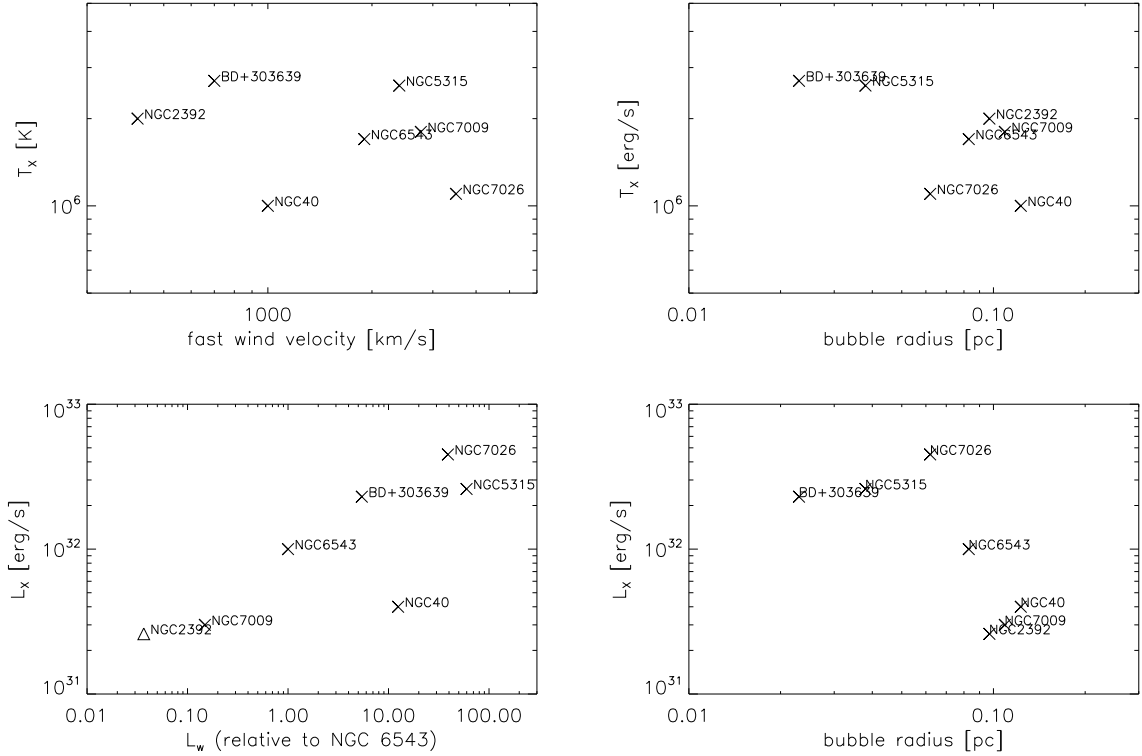}
\end{center}
\caption{ Results obtained thus far for PNe
  detected as ``hot bubble'' X-ray sources by Chandra and
  XMM-Newton (Table 1). Upper left:
  PN plasma temperature ($T_X$) vs.\ central star fast wind
  velocity. Upper right: $T_X$ vs.\ PN optical bubble
  radius. Lower left: PN X-ray luminosity ($L_X$) vs.\ wind
  luminosity of the PN central star ($L_w$) relative to
  $L_w$ of the central star of NGC 6543 ($L_w$ for NGC 2392
  [triangle] is an upper limit). Lower right: $L_X$ vs.\ PN optical bubble
  radius.  } 
\label{fig:PNXraytrends} \end{figure}

In Fig.~\ref{fig:PNXraytrends} we display scatter diagrams for various
of the observed quantities listed for the X-ray-detected PNe in Table
1. The top panels of Fig.~\ref{fig:PNXraytrends} demonstrate that the
characteristic temperature of the X-ray-emitting plasma is far lower
than expected, based on simple shock models, in almost all PNs in
which diffuse X-ray emission has been detected thus far (consistent
with the early X-ray CCD spectroscopic results reported for BD
$+30^\circ$3639 by Arnaud et al.\ 1996). The heat generated by stellar
wind shocks should produce a post-shock temperature $T = 2.5\times10^6
(v_w/400)^2$ (where $v_w$ is in km s$^{-1}$; e.g., Stute \& Sahai 2006
and references therein). There is only
one Table 1 object, NGC 2392, for which the temperature so predicted
is consistent with $T_X$; for all other PNe, the predicted post-shock
temperatures are larger than observed by factors ranging from $\sim2$
(BD $+30^\circ$3639) to $\sim200$ (NGC 7026).  Furthermore, $T_X$
appears uncorrelated with present-day central star wind velocity, but
is weakly anticorrelated with PN bubble radius
(Fig.~\ref{fig:PNXraytrends}, upper panels), suggesting that PN age is
more important than present-day wind kinetic energy in determining the
temperature of the X-ray-emitting plasma.

The bottom panels of Fig.~\ref{fig:PNXraytrends} indicate
that X-ray luminosity is correlated with present-day central
star wind luminosity $L_w = \frac{1}{2} \dot{M}v_w^2$, and
is perhaps anticorrelated with bubble radius. In the $L_X$ vs.\
$L_w$ plot (Fig.~\ref{fig:PNXraytrends}, lower left), NGC 40
appears somewhat underluminous in X-rays relative to the
other PNe, consistent with the ``punctured'' appearance of
its central bubble (Montez et al.\ 2005 and references
therein).
The shallow slope of the $L_X$ vs.\ $L_w$ correlation ---
wherein 4 orders of magnitude in $L_w$ results in only a
factor $\sim20$ range in $L_X$ --- may suggest either that
the conversion of wind kinetic energy to plasma radiation
becomes more efficient as the central star wind declines in
strength, or that the luminosities (and, perhaps, temperatures) of
the hot bubbles within these PNe are
established during early phases of the post-AGB evolution
of central stars with rapidly evolving winds 
(Akashi et al.\ 2006, 2007). 

\section{Conclusions}

We have detected luminous ($L_X \sim2.5\times10^{32}$ erg s$^{-1}$),
soft ($T_X \sim2.5\times10^{6}$ K) X-ray emission from the [WC] PN
NGC~5315. The emission most likely emanates from a relatively compact
($\sim8000$ AU radius), wind-blown bubble within this PN. Placed in
the context of results obtained to date by {\it Chandra} and {\it
  XMM}, these results for NGC~5315 indicate that the combination of
(1) relatively young, massive PNe harboring central stars with large wind
kinetic energies ($L_w \stackrel{>}{\sim} 10^{33}$ erg s$^{-1}$) and
(2) a ``closed containment vessel'' is necessary to yield PN hot
bubbles with plasma densities sufficient to produce detectable soft
(0.3-2.0 keV) X-ray luminosities $L_X \stackrel{>}{\sim} 10^{31}$ erg
s$^{-1}$. There should exist many other examples of diffuse, ``hot
bubble'' X-ray emission in addition to those listed in Table 1,
particularly among [WC] PNe with large ``present-day'' central star
wind luminosities and ``closed'' bubble/lobe morphologies. A
systematic {\it Chandra} and {\it XMM} survey of such PNe as well as a
set of ``control'' objects --- in particular, [WC] PNe with
open/amorphous morphologies on the one hand, and PNe with low-$L_w$
central stars but closed, bubble-like structures on the other ---
would test these assertions. In addition, to determine the longevity
of X-ray-luminous hot bubbles, it is important to search for X-ray
emission from the descendants of [WC] PNe, i.e., PNe with PG~1159-type
central stars. In this regard, it is intriguing that two such objects,
K 1--16 and NGC 246 (Jeffery et al.\ 1996), apparently have gone
undetected by {\it Chandra}\footnote{Based on preliminary analysis of
  archival data; see http://www.iaa.es/xpn/.}.

Collectively, these results provide further motivation for
ongoing modeling efforts that attempt to incorporate
processes such as the early onset (and rapid
decline) of the central star's fast wind, adiabatic cooling
of the (expanding) hot bubble, and heat conduction between
the shocked wind and nebular gas in predicting the range and
evolution of $L_X$ and $T_X$ in PNe (see discussions in
Soker \& Kastner 2003; Akashi et al.\ 2006, 2007; Stute \&
Sahai 2006; Sch\"onberner et al.\ 2006). Additional,
detailed modeling of a much larger and more representative
database of X-ray observations of PNe will be required to
establish which (if any) of the above processes are
particularly significant in determining $L_X$ and $T_X$
within PN hot bubbles. 

\acknowledgments
This research was supported by NASA through Chandra awards
GO4--5169X and GO5--6008X issued to Rochester Institute of
Technology by the Chandra X-ray Observatory Center, which is
operated by Smithsonian Astrophysical Observatory for and on
behalf of NASA under contract NAS8--03060. The authors wish to
acknowledge important contributions to (and comments on) this paper by Noam
Soker. 

\facility{CXO(ACIS)}




\clearpage

\begin{deluxetable}{cccccccc}
\rotate
\tablecolumns{8}
\tabletypesize{\small}
\tablecaption{Properties of Diffuse X-ray Planetary
  Nebulae\tablenotemark{a}\label{tab:objtable}}  
\tablehead{
\colhead{} &
\colhead{} &
\colhead{$D$} &
\colhead{$V_w$} &
\colhead{$dM/dt$} &
\colhead{$R_B$\tablenotemark{b}} &
\colhead{$L_X$} &
\colhead{$T_X$} \\
\colhead{Object} &
\colhead{Sp.\ type} &
\colhead{(kpc)} &
\colhead{(km s$^{-1}$)} &
\colhead{($M_\odot$ yr$^{-1}$)} &
\colhead{(pc)} &
\colhead{($10^{32}$ ergs s$^{-1}$)} &
\colhead{($10^6$ K)} 
}
\startdata
NGC 7026 & [WO 3] (1) & 1.7 (3) & 3500 (6) & $4.6\times10^{-7}$ (6) & 
  0.062 & $4.5$ (7) & $1.1$ (7) \\
NGC~5315 & [WC 4] (1) & 2.5 (4) & 2400 (4) & $1.5\times10^{-6}$ (4) & 
  0.038 & $2.6$ (8) & $2.6$ (8) \\
BD $+30^\circ3639$ & [WC 9] (1) & 1.2 (4) & 700 (4) & $1.6\times10^{-6}$ (4) &
  0.023 & $2.3$\tablenotemark{c} (9) & $2.2$ (10) \\
NGC 6543 & Of/WR(H) (2) & 1.8 (2) & 1900 (2) & $4.0\times10^{-8}$ (2) &
  0.083 & $1.0$ (11) & $1.7$  (11) \\
NGC 40 & [WC 8] (1) & 1.4 (4) & 1000 (4) & $1.8\times10^{-6}$ (4) & 
  0.12 & $0.40$\tablenotemark{d} (12) & $1.0$ (12) \\
NGC 7009 & O(H) (2) & 1.6 (2) & 2770 (2) & $2.8\times10^{-9}$ (2) & 
  0.11 & $0.30$ (13) & $1.8$ (13) \\
NGC 2392 & Of(H) (2) & 0.9 (2) & 420 (2) & $<3.0\times10^{-8}$ (2) & 
  0.10 & $0.26$ (14) & $2.0$ (14) \\
Hen 2-99 & [WC 9] (1) & 2.5 (5) & 900 (5) & $2.6\times10^{-6}$ (5) & 
  \nodata\tablenotemark{e} & $<0.05$ (12) & \nodata \\
\enddata

\tablenotetext{a}{Does not include the PNe NGC 7027, Menzel
  3, and Hen 3-1475, whose X-ray-emitting regions are
  likely generated by collimated winds or jets
  (Kastner et al.\ 2001, 2003; Sahai et al.\ 2003). A hot
  bubble was also detected within NGC 3242 by XMM (see
  http://www.iaa.es/xpn/), but results for $L_X$ and $T_X$
  are not yet available. References for spectral type of PN central
  star, adopted PN distance ($D$), PN central star wind
  velocity ($V_w$) and mass loss rate ($dM/dt$), and PN X-ray luminosity
  ($L_X$) and temperature ($T_X$) are listed in
  parentheses.}  

\tablenotetext{b}{Radius of central bubble, based on data reported in Cahn
  et al.\ 1992. In most cases, values of $R_B$ have been rescaled
  based on the adopted PN distances in column 3.}  

\tablenotetext{c}{$L_X$ for BD
  $+30^\circ3639$ has been rescaled based on the revised
  distance estimate of 1.2 kpc. }

\tablenotetext{d}{$L_X$ for NGC 40 has been
  revised upwards from that reported in Montez et al.\
  (2005) as a consequence of model fitting of reprocessed
  data; the fit results indicate that we had previously not properly
  accounted for ACIS contamination
  (http://cxc.harvard.edu/ciao/threads/aciscontam/). }

\tablenotetext{e}{In ground-based optical imaging, 
  Hen 2-99 displays a compact core and
  $\sim10''$ radius halo (Montez et al. 2005).}

\tablerefs
{1. Crowther et al. 1998; 2. Tinkler \& Lamers 2002 and
  references therein; 3. Hyung \& Feibelman 2004; 4. Marcolino et al.\ 2007; 
  5. Leuenhagen et al. 1996; 6. Koesterke \& Hamann 1997;
  7. Gruendl et al.\ 2006; 8. this paper; 9. Kastner et al.\ 2000; 
  10. Kastner et al.\ 2006; 11. Chu et al.\ 2001; 12. Montez et al.\ 2005;
  13. Guerrero et al.\ 2002; 14. Guerrero et al.\ 2005.
}

\end{deluxetable}

\clearpage

\end{document}